\documentclass[a4paper,11pt]{elsarticle}
\usepackage[utf8]{inputenc}
\usepackage[margin=1in]{geometry}
\usepackage{amsmath}
\usepackage{amssymb}
\usepackage{graphicx}
\usepackage{xcolor}
\usepackage[caption=false]{subfig}

\title{Global monopoles in the two-Higgs-doublet-model}
\author{Richard A. Battye, Steven J. Cotterill and Dominic G. Viatic}
\address{Jodrell Bank Centre for Astrophysics, Department of Physics and Astronomy, University of Manchester, Oxford Road, Manchester M13 9PL}

\begin{document}

\begin{abstract}
We discuss monopoles formed due to the spontaneous breakdown of a global $SO(3)_{\rm HF}$ symmetry within the global two-Higgs doublet model. We explain that the Higgs sector dynamics can be described in terms of two vectors one of which is null, $R^A=(R^0,R^a,R^4,R^5)$ for $a=1,2,3$, with 5 independent components describing the Higgs family symmetry and another, $n^a$, with 3 independent components related to the ``would-be'' Goldstone bosons. When formed from random initial conditions we find that monopoles are formed with a charged vacuum in the centre which couples the two fields together. We find a spherical symmetric solution which is an approximately uniform, unit winding of the sphere in both the $R^a$ and $n^a$ vectors. These global monopoles are closely related to the Nambu monopole. The additional complexity and structure contained in these monopoles does not appear to prevent the scaling of their density.
\end{abstract}

\maketitle

\section{Introduction}
    \label{sec:intro}

Monopoles are topological defects which arise from the breaking of a spherical symmetry~\cite{Shellard1994} and can be formed during the breaking of both \textit{global} or \textit{local/gauge} symmetries. Gauge monopoles are commonly predicted in GUTs~\cite{tHooft1974} as they typically occur in phase transitions where the ``little group'' contains a U(1) symmetry. As with domain walls, gauge monopoles can present issues for late-time cosmology (see, for example, ref.~\cite{Martins2008}). The energy density due to monopoles  is expected to scale like matter in an FRW universe, and the ratio of the two will remain constant. If the initial density of monopoles is larger than that of matter, it will remain so throughout cosmic history preventing a matter dominated epoch. 

A commonly suggested solution to this issue is a period of cosmic inflation in the early Universe which could dilute the initial energy density of monopoles to a level that would be compatible with the fact that we have not observed them (see, for example, ~\cite{Burdin:2014xma}). However, the global monopoles considered in this article are predicted to emerge at the electroweak scale and are, therefore, expected to be post-inflation monopoles. 


The Higgs mechanism for electroweak symmetry breaking was verified by the measurement of a Higgs boson~\cite{Aad2012,Chatrchyan2012} of mass 125~GeV at the LHC~\cite{Aad2015}. The properties of this scalar particle so far match those predicted for the Standard Model (SM) Higgs scalar~\cite{Djouadi:2005gi,Khachatryan2016}.
Nonetheless, current experimental measurements do not prohibit the existence of more scalar particles.
One minimal and theoretically well-motivated extension which can be made to the SM is to introduce a second complex Higgs doublet into the theory. This is the so-called \textit{two-Higgs-doublet-model} (2HDM) \cite{Lee1973,Pilaftsis1999,Branco2012}. 

The 2HDM allows for the emergence of a variety of topological defects, such as domain walls, vortices and global monopoles, from the breaking of accidental symmetries which the model can possess under certain parameter choices ~\cite{Brawn2011,Eto:2018hhg,Eto:2018tnk,Chen:2020soj,Viatic2020,Viatic2020b,Eto:2020opf,Eto:2020hjb,Eto:2021dca,Law:2021ing}. Phase transitions associated with the spontaneous breaking of these symmetries can leave relic topological defects which can serve as probes of high energy physics in the early Universe~\cite{Garagounis2003,Kibble:1982dd,Nakayama2017}.
In this letter we will focus our attention on the $\text{SO}(3)_\text{HF}$-symmetric 2HDM which predicts global monopoles~\cite{Brawn2011}. The key question we want to answer is whether the phenomena found in ref.~\cite{Viatic2020}, whereby a massive photon (what is often termed a ``charged vacuum") is predicted at the centre of the domain wall when the defects are formed from random initial conditions, is a generic one for other topological defects in the 2HDM.

We will ignore the SM gauge degrees of freedom in our simulations since these make the simulations technically more difficult. The introduction of gauge fields will introduce new length-scales into the problem which would need to be carefully dealt with in the simulations and it is our strong belief, based on experience of the evolution of topological defects in other contexts, that is unlikely to have a significant impact on the qualitative picture of the topological defects produced.


\section{Two-Higgs-doublet-model with $\text{SO}(3)_\text{HF}$ symmetry}
\label{sec:2hdm}

The Lagrangian density for the model is 
\begin{equation}\label{eq:2HDMlagrangian}
    \mathcal{L} =  \sum_{i=1}^{2} (\partial^\mu \Phi_i)^\dagger (\partial_\mu \Phi_i) - V(\Phi_1,\Phi_2)\,,
\end{equation}
where $\Phi_1$ and $\Phi_2$ are two complex doublet fields, and the potential is given by
\begin{equation}\label{eq:SO3Potential}
    \begin{split}
    V(\Phi_1,\Phi_2) = & -\mu_1^2 \left(\Phi_1^\dagger\Phi_1 + \Phi_2^\dagger\Phi_2\right) + \lambda_1\left( \Phi_1^\dagger\Phi_1 + \Phi_2^\dagger\Phi_2\right)^2 \\
    & + \lambda_4\left[(\Phi_1^\dagger\Phi_2)(\Phi_2^\dagger\Phi_1) - (\Phi_1^\dagger\Phi_1)(\Phi_2^\dagger\Phi_2)\right]\,.
    \end{split}
\end{equation}
This potential possesses a $\text{SO}(3)_\text{HF}\cong SU(2)_\text{HF}/Z_2$ symmetry and has 3 real parameters; $\mu_{1}^2, \lambda_1$ and $\lambda_4$\footnote{The parameters $\mu_1$, $\lambda_1$ and $\lambda_4$ have been chosen to fit with standard notation eg. \cite{Brawn2011}.}, where $\lambda_4$ prevents the potential from being symmetric under a larger symmetry group, but is also crucial for preventing neutral vacuum violation \cite{Brawn2011}.
Under a $\text{SO}(3)_\text{HF}$ transformation the complex scalar Higgs doublets, $\Phi_1$ and $\Phi_2$, transform as $\Phi_1 \rightarrow e^{-i\alpha} c_\gamma \Phi_1 + e^{-i\beta} s_\gamma\Phi_2$, $\Phi_2 \rightarrow - e^{i\beta} s_\gamma \Phi_1 + e^{i\alpha} c_\gamma \Phi_2$,
where $0\le\alpha,\beta,\gamma<\pi$ with $s_\gamma=\sin\gamma$ and $c_\gamma=\cos\gamma$, that is, an $SO(3)$ Higgs Family (HF) rotation between the two Higgs doublets which is broken to $U(1)$ and hence the vacuum manifold contains an extra $S^2$. This is an additional symmetry on top of those associated with the SM, see ref.~\cite{Brawn2011} for a detailed discussion.

The most general parametrization of the 2HDM vacuum is \cite{Branco2012}
\begin{equation}\label{eq:ChargedVac}
{\bar\Phi}_1 = \frac{1}{\sqrt{2}}\left(\begin{matrix}
0 \\
v_1
\end{matrix}\right),\quad 
{\bar\Phi}_2 = \frac{1}{\sqrt{2}}\left(\begin{matrix}
v_+ \\
v_2 e^{i\xi}
\end{matrix}\right)\,.
\end{equation}
In the following section we will use $ v_1,v_2,v_+ $ and $ \xi $ to parameterise the fields. However, for the moment let us think of them as just the vacuum and expand around a neutral vacuum with $v_+=0$ and $\xi=0$. The 2HDM has 5 physical scalar particles: 2 neutral CP-even states, $h$ and $H$, one CP-odd neutral state, $A$, and 2 charged states, $H^\pm$.
The other three scalar degrees of freedom correspond to would-be Goldstone bosons, $G^0$ and $G^\pm$, which are absorbed into the longitudinal components of the electroweak gauge bosons, $W^\pm$ and $Z^0$.
Expressions for the masses of the scalar Higgs particles, $h$, $H$, $A$ and $H^\pm$, are obtained as eigenvalues of the Hessian matrix of \eqref{eq:SO3Potential} using the parametrization
\begin{equation}\label{eq:PWrep}
    \Phi_1 = \left(\begin{matrix}
    \varphi_1^+ \\
    \frac{1}{\sqrt{2}}v_1 + \varphi_1 + i a_1
    \end{matrix}\right),\quad
    \Phi_2 = e^{i\xi}\left(\begin{matrix}
    \varphi_2^+ \\
    \frac{1}{\sqrt{2}}v_2 + \varphi_2 + i a_2
    \end{matrix}\right)\,,
\end{equation}
where $\varphi_i^+$ are complex scalar fields.
One finds that, due to the greatly reduced parameter space required to obtain $SO(3)_\text{HF}$ symmetry, one of the CP-even scalars along with the CP-odd scalar are both massless,
$M_H = 0, M_A = 0$ and the other CP even scalar is the SM Higgs with $M_h^2 = 2\lambda_1v_\text{SM}^2$.
The charged scalar mass is given by $ M_{H^\pm}^2 = -\frac{1}{2}\lambda_4 v_\text{SM}^2.$
Therefore, to ensure a stable vacuum, the quartic coupling parameters must satisfy the inequalities,
$\lambda_1 > 0,\lambda_4 < 0$,
such that the non-zero scalar masses are real and positive. The fact that there are massless Higgs states would probably make this specific manifestation of the model not phenomenologically viable, but many of the features found here are likely to persist in more realistic models.

\section{Field parameterization}

\begin{figure}
    \centering
    \includegraphics[width=\textwidth]{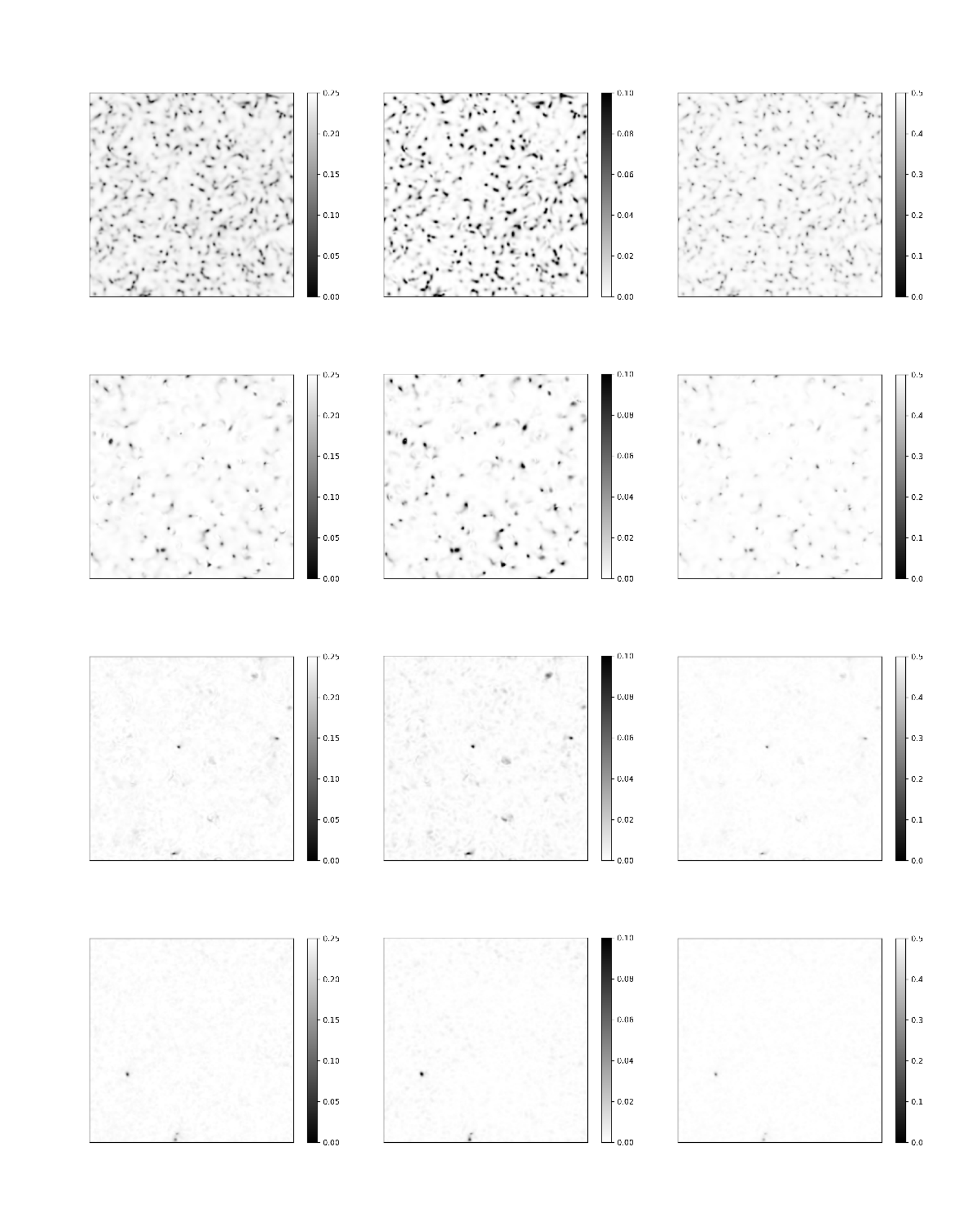}
    \caption{Two-dimensional slices showing spatial distributions of $\sqrt{n^a n^a}$ (left), $R^\mu R_\mu$ (middle) and $R^a R^a$ (right) from a (3+1) dimensional simulation of the $\text{SO}(3)_\text{HF}$-symmetric 2HDM with $\epsilon= 0.35$. Note that the colour bars have been chosen so that all three quantities look very similar. The simulation with $P=512$, $\Delta t=0.2$ and $\Delta x=0.9$  was run for a time $t=240$ and dissipation was removed at $t=30$. Time increases downwards with the top plot being $t=30$, next two $t=60,120$ and the bottom plot corresponds to the end of the simulation, $t=240$  The global monopoles are located at the point where $R^aR^a=0$ which correlates strongly with maxima in $R^\mu R_\mu$ indicating that charge neutrality condition is violated at the core of the monopoles. Away from the monopoles $\sqrt{n^an^a}=1/2$ and this quantity takes its minimum value at the monopole cores. As time evolves the density of monopoles can been seen to visually reduce roughly scaling like $t^{-3}$. See later for further discussion of the scaling exponent.}
    \label{fig:2Dmonoslice}
\end{figure}

It turns out, that in terms of understanding the topological defect solutions, the doublet fields $\Phi_1$ and $\Phi_2$ are not the best way to see the structure of the vacuum. This is because the global symmetries which are being spontaneously broken are due to the internal symmetries between the components of the doublets. It was explained in ref.\cite{Brawn2011} that the additional degrees of freedom found in the 2HDM can be parameterized in terms of a vector $R^A$ with $A=0,1,..,5$ which is null in the sense that $R^AR_A=0$ and hence there are 5 degrees of freedom. These can be thought of as  corresponding to the 5 Higgs particles in the 2HDM. For $A=\mu=0,1,.,3$ we have that~\cite{Maniatis2007,Ivanov2007,Nishi:2006tg,Brawn2011} \begin{equation}\label{eq:RVector}
	R^\mu = \Phi^\dagger \left(\sigma^\mu \otimes I_2\right) \Phi = \left(\begin{matrix}
	\Phi_1^\dagger\Phi_1 + \Phi_2^\dagger\Phi_2 \\
	\Phi_1^\dagger\Phi_2 + \Phi_2^\dagger\Phi_1 \\
	-i[\Phi_1^\dagger\Phi_2 - \Phi_2^\dagger\Phi_1] \\
	\Phi_1^\dagger\Phi_1 - \Phi_2^\dagger\Phi_2
	\end{matrix}\right)\,,
\end{equation}
where $\sigma^\mu$ are the Pauli matrices including the identity, $\sigma^0 = I_2$ and we have introduced the multiplet,
\begin{equation}
\Phi \equiv\left(\begin{matrix}\Phi_1\cr\Phi_2\end{matrix}\right)\,.
\end{equation}
The other two components are 
\begin{equation}
    R^4=\Phi_1^Ti\sigma^2\Phi_2-\Phi_2^{\dagger}i\sigma^2\Phi_1^{*}\,,\quad R^5=-i\left( \Phi_1^{T}i\sigma^2\Phi_2+\Phi_2^{\dag}i\sigma^2\Phi_1^*\right)\,,
\end{equation}
which can be brought together in a complex field $\tilde{R}=R^4+iR^5=2\Phi^T_1i\sigma^2\Phi_2$. In a neutral vacuum, where the photon mass is zero, we have that $v_+=0$ and hence we find that $R^\mu R_\mu=0$ and $\tilde{R}=0$. It has already been established~\cite{Viatic2020} that simulations of 2HDM domain walls predict a violation of this neutral vacuum condition in the core of the defect as a general feature emerging from random initial conditions and we also expect this to be the case for the global monopoles considered here.

Now define the 3-component field,
\begin{equation}
    n^a = -\Phi^\dagger (I_2 \otimes \sigma^a) \Phi = -\sum_{n=1}^2 \Phi_n^\dagger \sigma^a \Phi_n\,,
\end{equation}
for $a=1,2,3$. We will see that this encodes the 3 degrees of freedom which will be ``eaten" by the gauge fields as part of the Higgs mechanism.

We can parameterise the field as $\Phi=e^{i\chi/2}(I_2\otimes U_L){\bar\Phi}$ where $0\le\chi<2\pi$, $U_L$ is an element of $SU(2)_L$ and ${\bar\Phi}={1\over\sqrt{2}}\left(0,v_1,v_+,v_2e^{i\xi}\right)$, that is, a general element of the vacuum manifold. There are a total of eight degrees of freedom $v_1,v_2,v_+,\xi,\chi$ and the three degrees of freedom contained within $U_L$. Immediately we see that $R^{\mu}={\bar\Phi}^{\dag}(\sigma^{\mu}\otimes I_2){\bar\Phi}={\bar R}^{\mu}$ with
\begin{equation}
{\bar R}^\mu= {1\over 2}\left(\begin{matrix}
	v_1^2+v_2^2+v_+^2 \\
	2v_1v_2\cos\xi \\
	2v_1v_2\sin\xi \\
	v_1^2-v_2^2-v_+^2
	\end{matrix}\right)\,,
\end{equation}
which only depends on $v_1,v_2,v_+$ and $\xi$, and is independent of $\chi$ and $U_L$.  We also see that $\tilde{R}=2e^{i\chi}{\bar\Phi}_1^Ti\sigma^2{\bar\Phi}_2=e^{i\chi}\tilde{\bar R}$ where $\tilde{\bar R}=-v_+v_1$. This is also independent of $U_L$ as a consequence of $U_L^Ti\sigma^2U_L=i\sigma^2$ and this illustrates that the phase of the complex scalar field, ${\tilde R}$, isolates $\chi$. Therefore, $R^\mu$ and $\tilde{R}$ encode the four new degrees of freedom added in going from the SM to the 2HDM as well as the SM Higgs.

Under the same parameterization we see that
\begin{equation}
    n^a=-{\bar\Phi}^{\dagger}\left(I_2\otimes (U_L^{\dagger}\sigma^aU_L)\right){\bar\Phi}\,,
\end{equation}
which is independent of $\chi$. If we write $U_L=\exp\left[iG^a\sigma^a/(2v_{\rm SM}\right)]$ where the $G^a$ are the three ``would be" Goldstone bosons, then $U_L^{\dagger}\sigma^aU_L={\cal R}^{ab}\sigma^b$ where 
\begin{equation}
{\cal R}^{ab}={\hat G}^{a}{\hat G}^{b}+(\delta^{ab}-{\hat G}^{a}{\hat G}^{b})\cos \gamma+\epsilon^{abc}{\hat G}^c\sin \gamma\,,
\end{equation}
$\gamma=|G|/v_{\rm SM}$ and ${\hat G}^a=G^{a}/|G|$, that is, a rotation by an angle $\gamma$ around the unit vector ${\hat G}^a$. Hence, we can write $n^{a}={\cal R}^{ab}{\bar n}^b$ where ${\bar n}^{a}=-{\bar\Phi}^{\dagger}(I_2\otimes\sigma^a){\bar\Phi}$ and in a neutral vacuum state we have $\bar{n}^a=\textstyle{1\over 2}v_{\rm SM}^2(0,0,1)$. This shows that action of the SM degrees of freedom leads to a rotation of ${\bar n}^a$, but remember that they have no impact on $R^{\mu}$.

\section{Formation and evolution of global monopoles}
    \label{sec:sims}

\begin{figure}
    \centering
    \vspace{-0.5cm}
    \subfloat{
        \centering
        \includegraphics[trim={4.5cm 2.5cm 4.5cm 5.5cm},clip,width=0.48\linewidth]{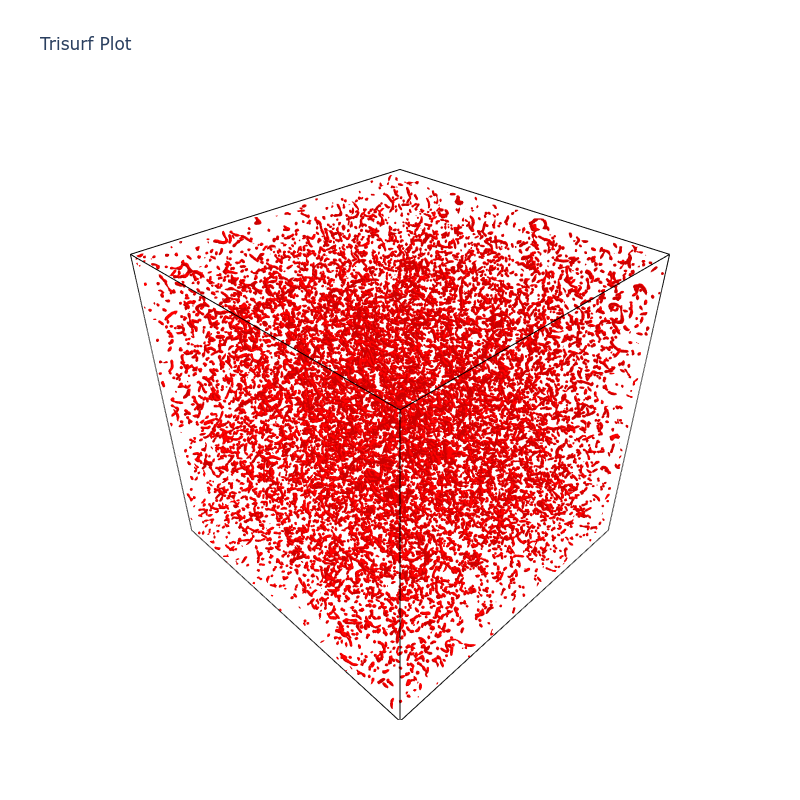}
    }
    \vspace{-0.5cm}
    \subfloat{
        \centering
        \includegraphics[trim={4.5cm 2.5cm 4.5cm 5.5cm},clip,width=0.48\linewidth]{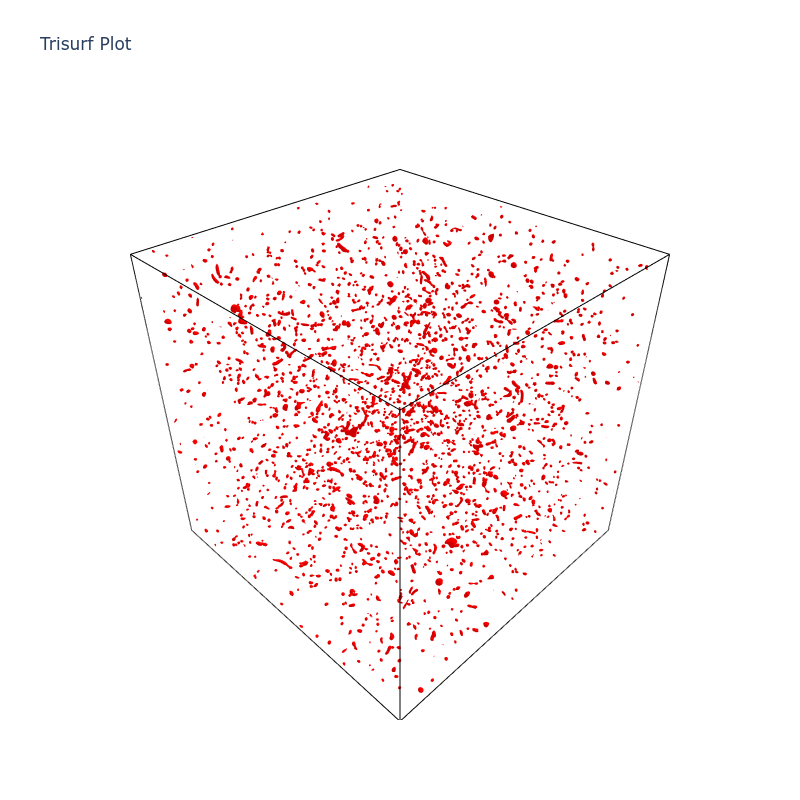}
    }\\
    \subfloat{
        \centering
        \includegraphics[trim={4.5cm 2.5cm 4.5cm 5.5cm},clip,width=0.48\linewidth]{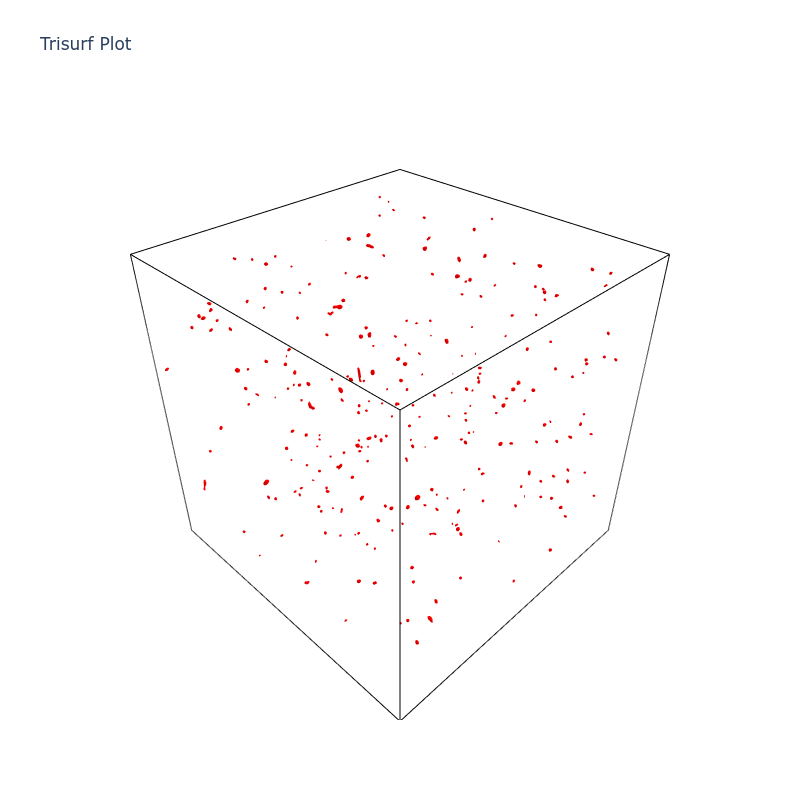}
    }
    \subfloat{
        \centering
        \includegraphics[trim={4.5cm 2.5cm 4.5cm 5.5cm},clip,width=0.48\linewidth]{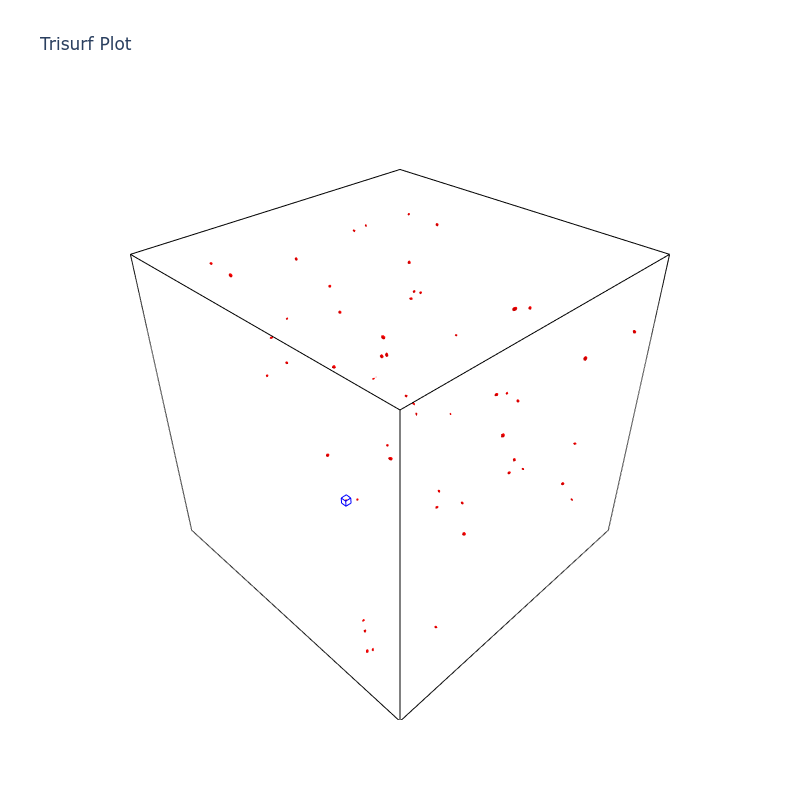}
    }\\
    \includegraphics[width=0.6\textwidth]{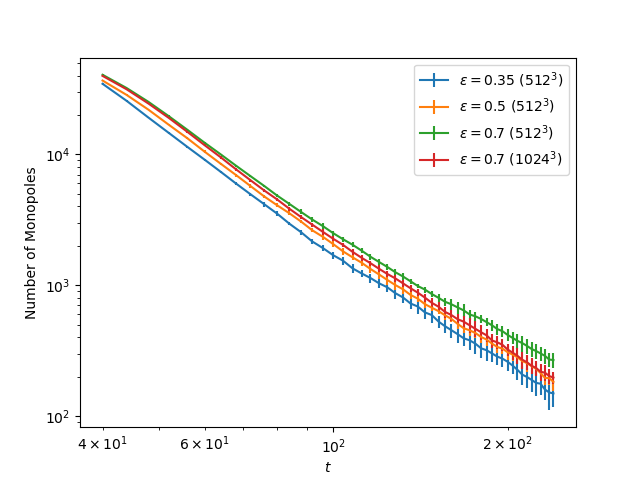}
    \caption{(top) Isosurfaces of $R^{\mu}R_{\mu}=0.1$ - a proxy for the position of the monopole - as a function of time for $\epsilon=0.35$. The times used are $t=30$ (top-left), $t=60$ (top-right), $t=120$ (middle-left) and $t=240$ (middle-right) as in Fig.~\ref{fig:2Dmonoslice}. In the $t=240$ plot there is a small blue box which isolates a small region used in Fig.~\ref{fig:fieldzoom}. At the bottom we show the evolution of the number of monopoles, for various values of $\epsilon$, averaged over 10 realisations as described in the text. Error bars represent the numerical scatter via the standard deviation.}
    \label{fig:3Diso}
\end{figure}


We have performed (3+1) dimensional simulations for the global scalar field theory of the 2HDM with $\text{SO}(3)_\text{HF}$ symmetry by evolving the equations of motion on a regular grid of $P^3$ points for $P=512$ and $P=1024$ with Minkowski metric and periodic boundary conditions (for details of the simulation procedure, see \cite{Viatic2020,Viatic2020b}).
Temporal derivatives are approximated to second order and spatial derivatives to fourth order. The initial conditions are created by placing small random numbers in all of the fields and then evolving them under dissipative dynamics for around 200 time steps, before allowing the fields to evolve under the full dynamics up to the light crossing time set by $P\Delta x/2$ where $\Delta x=0.9$ when $P=512$ and $\Delta x=0.45$ when $P=1024$ is the spatial stepsize. We use a timestep of $\Delta t=0.2$ for $P=512$ and $0.1$ for $P=1024$ in order to maintain numerical stability.

The units can be rescaled so the energy density is measured in units of $\mu_1^4/\lambda_1$ and length/time in units of $\mu_1^{-1}$. This leaves a single dimensionless parameter $\epsilon=M_{H^{\pm}}/M_h$. We ran simulations for three  different choices of $\epsilon=0.35,0.5$ and $0.7$, with the largest corresponding to $M_{H\pm}\approx 88\,{\rm GeV}$. This is quite low for the charged Higgs mass, but having a low value of $\epsilon$ makes the simulation more numerically tractable. We expect the qualitative features of the simulations to be similar over a wide range of parameters, and in particular for much larger values of $\epsilon$, but this will be confirmed in future work.

In Fig.~\ref{fig:2Dmonoslice} we present two dimensional slices of the spatial distribution of the scalar quantities $\sqrt{n^a n^a}$, $R^\mu R_\mu$ and $R^a R^a$ for the simulation with $\epsilon=0.35$. The centres of monopoles are defined to be when $R^aR^a=0$. The right hand column indicates that monopoles form during the period of dissipation and that their density is initially high but that it rapidly decreases with time. The quantity $R^\mu R_\mu$ is non-zero at the cores of the monopoles and the maxima of this quantity are strongly correlated with the positions of the monopoles. This indicates that inside the monopoles the vacuum is charged with $v_+\ne 0$, which would correspond to a non-zero photon mass in the centre of the monopole. We also find that $\sqrt{n^a n^a}$ has its minimum value at the same place as the position of the monopoles. 

The 3D dynamics of the monopoles for $\epsilon=0.35$ are illustrated in Fig.~\ref{fig:3Diso} where we have plotted isosurfaces of the quantity $R^{\mu}R_{\mu}$ which, based on the previous paragraph, we now take as a proxy for the positions of the monopoles. Complex dynamics takes place, presumably with the monopoles interacting with each other and  annihilating with anti-monopoles. Close examination of the simulation output appears to reveal a number of dumb-bell-shaped configurations which could be monopoles/anti-monopoles interacting via a bridge comprising a string-like structure, motivating detailed investigation of the simulations that is beyond the scope of the present work. Similar objects have been examined in the context of the Standard Model~\cite{Lazarides:2021bzg,Patel:2023sfm} and $SU(5)$ Grand Unified Theories~\cite{PhysRevD.103.095021}. 

We also present the number of monopoles, $N_\text{mono}$, as a function of time obtained as an average over 10 realizations and we give the best fit scaling exponents in Table~\ref{tab: scaling exponents}. We note that in the case of $\epsilon=0.7$ it was necessary to use a lower value of $\Delta x$ (keeping $P\Delta x$ constant) in order to resolve the length scale $\propto \epsilon^{-1}$ - simulations with the higher value of $\Delta x$ indicated some deviation from th expected scaling law which was restored by higher resolution. We find that monopoles in the $\text{SO}(3)_\text{HF}$-symmetric 2HDM scale as $N_{\rm mono}\propto t^{-3}$ which is compatible with what is seen to take place in the global $O(3)$ model~\cite{Yamaguchi:2001rf,Bennett:1990xy}. It appears, therefore, that whatever leads to the violation of the neutral vacuum inside the monopoles does not lead to a significant change in the expected evolution, at least within the dynamical range of these simulations. It would be interesting to perform simulations with a larger dynamic range in order to test this.

\begin{table}[t]
\centering
\begin{tabular}{c|c c c}
    $\epsilon$ & $0.35$ & $0.5$ & $0.7$  \\
    \hline
    Scaling Exponent & $-2.9\pm0.1$ & $-2.93\pm0.08$ & $-2.98\pm0.08$ \\
\end{tabular}
\caption{Measured scaling exponents for three different values of $\epsilon$. Those for $\epsilon=0.35$ and $0.5$ used $P=512$ and $\Delta x=0.9$, while that for $\epsilon=0.7$ used $P=1024$ and $\Delta x=0.45$. In all three cases we see that they are compatible with the number of monopoles scaling $\propto t^{-3}$.}
\label{tab: scaling exponents}
\end{table}


\begin{figure}[t] \label{fig:fieldzoom}
    \centering
    \subfloat[$R^a$]{
        \centering
        \includegraphics[trim={2.5cm 1cm 2.2cm 2cm},clip,width=0.48\linewidth]{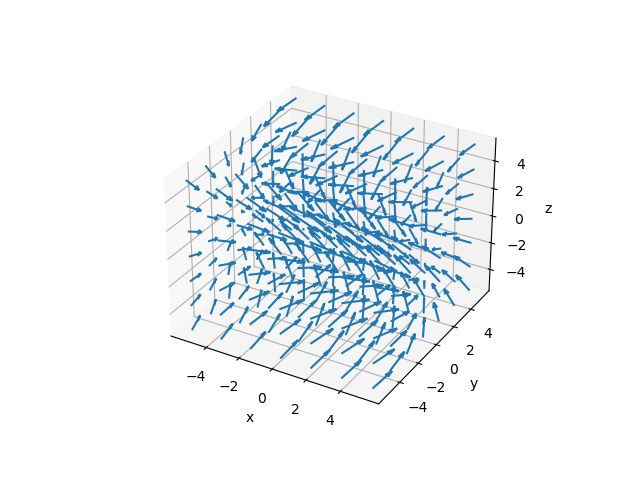}
    }
    \subfloat[$R^a$ after a global rotation]{
        \centering
        \includegraphics[trim={2.5cm 1cm 2.2cm 2cm},clip,width=0.48\linewidth]{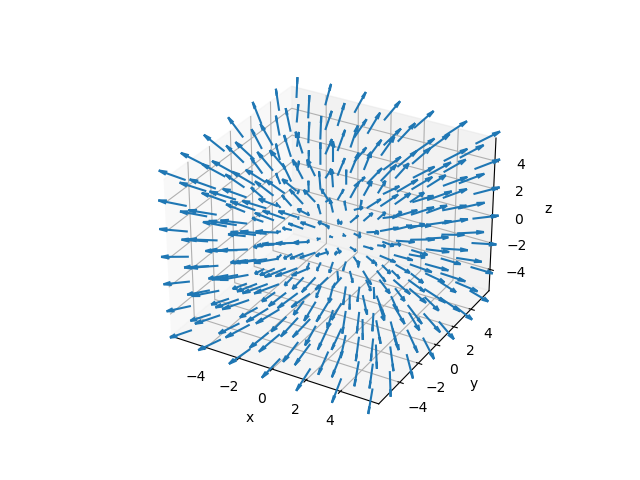}
    }\\
    \subfloat[$n^a$]{
        \centering
        \includegraphics[trim={2.5cm 1cm 2.2cm 2cm},clip,width=0.48\linewidth]{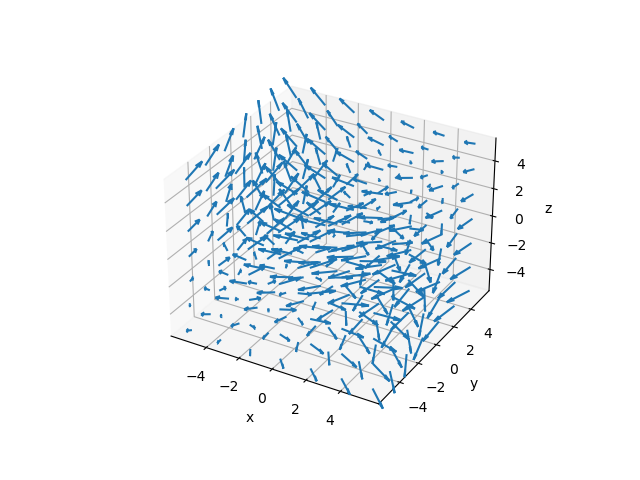}
    }
    \subfloat[$n^a$ after a global rotation]{
        \centering
        \includegraphics[trim={2.5cm 1cm 2.2cm 2cm},clip,width=0.48\linewidth]{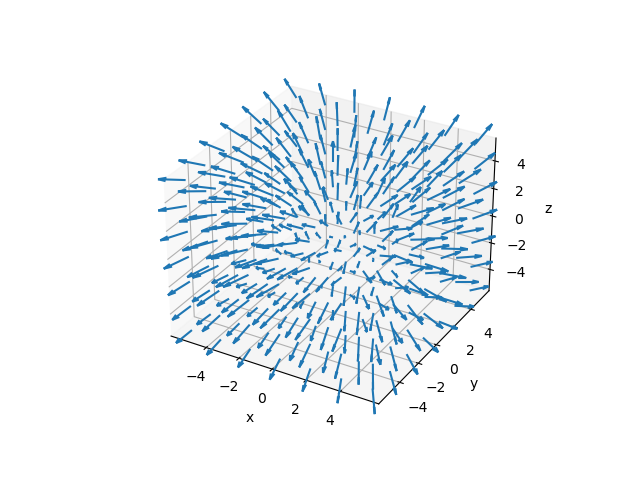}
    }
    \caption{ We present $R^a$ (top) and $n^a$ (bottom) around one of the monopoles at the end of the simulation. The left side shows the vectors that are directly calculation from the simulation and on the right we show the same vector fields after a global rotation has been applied. Note that it is a different rotation for $R^a$ and $n^a$. We have applied a similar procedure, but with different rotations, to some of the other monopoles in the simulation and are able to get similar results.}
\end{figure}

To further investigate the nature of the field configurations that are formed in the simulations, we have isolated a region at late times which appears to contain a single monopole. In Fig.~\ref{fig:fieldzoom} we have attempted within constraints of the resolution of the simulation, which only has a few grid squares for each monopole, to plot the vectors $R^{a}$ and $n^{a}$. At first glance the configurations indicate some spatial gradients, but are not immediately identifiable as monopoles. However, we have found that we can apply global rotations (not the same for both) to orient all of the vectors so that they point in a direction that is approximately radial in the vicinity of the monopole.



\section{Global monopole solution}

A global monopole solution was proposed~\cite{Brawn2011} which can written as 
\begin{equation}
\Phi={1\over\sqrt{2}}v_{\rm SM}f(r)\left({\hat {\bf r}}\cdot \sigma\otimes I_2\right)\left(\begin{matrix}0\cr 1\cr 0\cr 0\end{matrix}\right)\,,
\label{brawnmonopole}
\end{equation}
where $f(r)$ is a function just depending on the radial coordinate $r$ which satisfies a second order differential equation that can be deduced from ref.~\cite{Brawn2011} with boundary conditions $f(0)=0$ and $f(\infty)=1$ and $v_{\rm SM}=\mu_1/\sqrt{\lambda_1}$. It was constructed by setting $v_1=v_{\rm SM}f(r)\cos\theta$, $v_2=v_{\rm SM}f(r)\sin\theta$, $v_+=0$  and $\xi=\phi$.

This neutral vacuum solution has $R^{a}=\textstyle{1\over 2}v_{\rm SM}^2[f(r)]^2{\cal Q}^{ab}{\hat z}^b$ where ${\cal Q}^{ab}=2{\hat r}^a{\hat r}^b-\delta^{ab}$ and $n^a=\textstyle{1\over 2}v^2_{\rm SM}[f(r)]^2{\hat z}^a$, both of which are $\textstyle{1\over 2}v^2_{\rm SM}[f(r)]^2$ times a unit vector with $R^a={\cal Q}^{ab}n^b$. We now believe that this is in fact not a monopole since $R^a$ has zero winding number, and in fact is a sphaleron.

 In light of $R^a$ and $n^a$ that we have discovered in the monopoles formed from random initial conditions, we initially condsidered a monopole ansatz that has $R^a=n^a\propto \hat{r}^a$
\begin{equation} \label{eq: No charge-breaking monopole ansatz}
    \Phi = \frac{v_{\rm SM}f(r)}{2\sqrt{2}}\begin{pmatrix}
        -\sin\theta e^{-i\phi} \\
        \cos\theta + 1 \\
        \cos\theta - 1 \\
        \sin\theta e^{i\phi}
    \end{pmatrix} \,,
\end{equation}
which has $v_1 = v_{\rm SM}f(r)\cos\frac{1}{2}\theta$, $v_2 = v_{\rm sM}f(r)\sin\frac{1}{2}\theta$,  $v_+ = 0$ and $\xi=\phi$ with the $\text{SU(2)}_{\rm L}$ rotation
\begin{equation}
    U_L = \begin{pmatrix}
        \cos\frac{1}{2}\theta & -\sin\frac{1}{2}\theta e^{-i\phi} \\
        \sin\frac{1}{2}\theta e^{i\phi} & \cos\frac{1}{2}\theta
    \end{pmatrix} \,.
\end{equation}
The expressions for $v_1$ and $v_2$ are half angle versions of (\ref{brawnmonopole}) and $U_L$ generates a Nambu monopole in the SM~\cite{NAMBU1977505}, but in this case there is no need for a string to be attached to one of the poles. This is because the divergent part of the gradient energy associated with the winding of $n^a$ cancels due to the extra winding in $R^a$. However, this ansatz cannot describe the monopoles that we see in the simulations as $R_\mu R^\mu = 0$ (because $v_+=0$) and $\Phi$ will be forced to zero at the core of the monopole by the gradient energy terms. 

A simple extension that allows for charge-breaking in the core of the monopole is
\begin{equation}
    \Phi = \frac{v_{\rm SM}}{2\sqrt{2}}\begin{pmatrix}
        -(f-f_+)\sin\theta e^{-i\phi} \\
        (f-f_+)\cos\theta + (f+f_+) \\
        (f-f_+)\cos\theta - (f+f_+) \\
        (f-f_+)\sin\theta e^{i\phi}
    \end{pmatrix} \,,
\end{equation}
where $f=f(r)$ and $f_+=f_+(r)$ for which  $R^a = n^a = \frac{v_{\rm SM}^2}{2}(f^2-f_+^2)\hat{r}^a$. This ansatz can be written as $(I_2 \otimes U_L)\Bar{\Phi}$ (as any configuration in the 2HDM can) but it is a much simpler expression in this form. Note that fixing $f_+=0$ returns the ansatz of equation (\ref{eq: No charge-breaking monopole ansatz}) and the gradient energy does not force $\Phi = 0$ at the centre of the monopole - only $f=f_+$. Under this ansatz, $R_\mu R^\mu = v_{\rm SM}^4f^2f_+^2$ and therefore there will be a massive photon and neutral vacuum violation whenever both $f$ and $f_+$ are simultaneously non-zero.

One can calculate the energy to be 
\begin{equation}
    \hat{E} = \int r^2dr\bigg\{\frac{1}{2}\bigg(\frac{df}{dr}\bigg)^2 + \frac{1}{2}\bigg(\frac{df_+}{dr}\bigg)^2 + \frac{(f-f_+)^2}{2r^2} - \frac{1}{2}(f^2+f_+^2) + \frac{1}{4}(f^2+f_+^2)^2 + \epsilon^2f^2f_+^2 \bigg\}\,,
\end{equation}
where $E = 4\pi v_{\rm SM}^2\hat{E}/\mu_1$, $v_{\rm SM} = \mu_1/\sqrt{\lambda_1}$ and we have rescaled the length scale, $r \to r/\mu_1$, so that the solution only depends upon the single parameter, $\epsilon = M_{\rm H\pm}/M_h$. Such a solution must satisfy,
\begin{eqnarray}
    \frac{1}{r^2}\frac{d}{dr}\bigg(r^2\frac{df}{dr}\bigg) - \frac{f-f_+}{r^2} - \bigg[ f^2 + f_+^2 - 1 + 2\epsilon^2f_+^2 \bigg]f &=& 0 \,, \\
    \frac{1}{r^2}\frac{d}{dr}\bigg(r^2\frac{df_+}{dr}\bigg) + \frac{f-f_+}{r^2} - \bigg[ f^2 + f_+^2 - 1 + 2\epsilon^2f^2 \bigg]f_+ &=& 0 \,,
\end{eqnarray}
with the boundary conditions $f(0)=f_+(0)$, $f'(0) = -f_+'(0)$, $f(\infty) = 1$ and $f_+(\infty)=0$.

\begin{figure}[t]
    \centering
    \subfloat{
        \centering
        \includegraphics[trim={0cm 0.5cm 1.5cm 1.8cm},clip,width=0.48\linewidth]{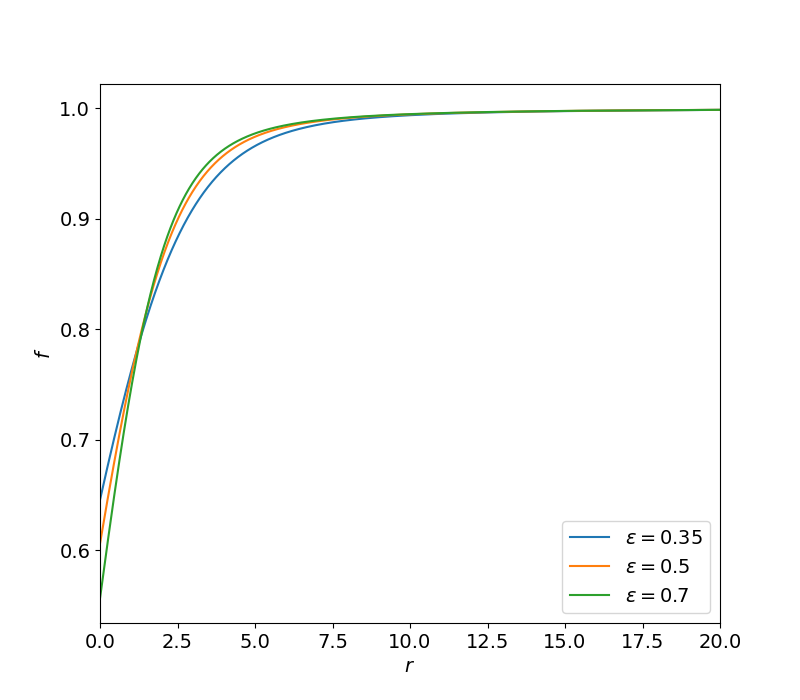}
    } 
    \subfloat{
        \centering
        \includegraphics[trim={0cm 0.5cm 1.5cm 1.8cm},clip,width=0.48\linewidth]{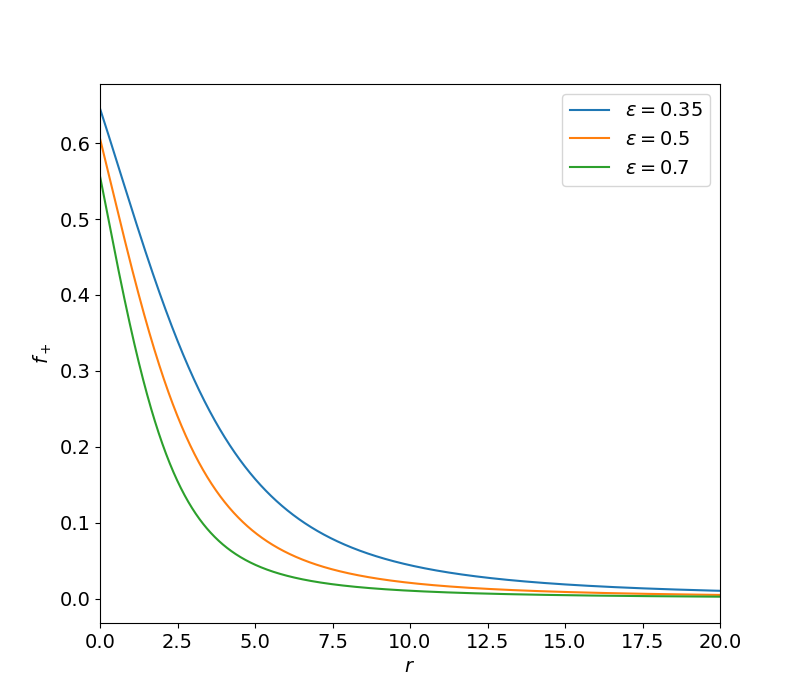}
    }\\
    \subfloat{
        \centering
        \includegraphics[trim={0cm 0.5cm 1.5cm 1.8cm},clip,width=0.48\linewidth]{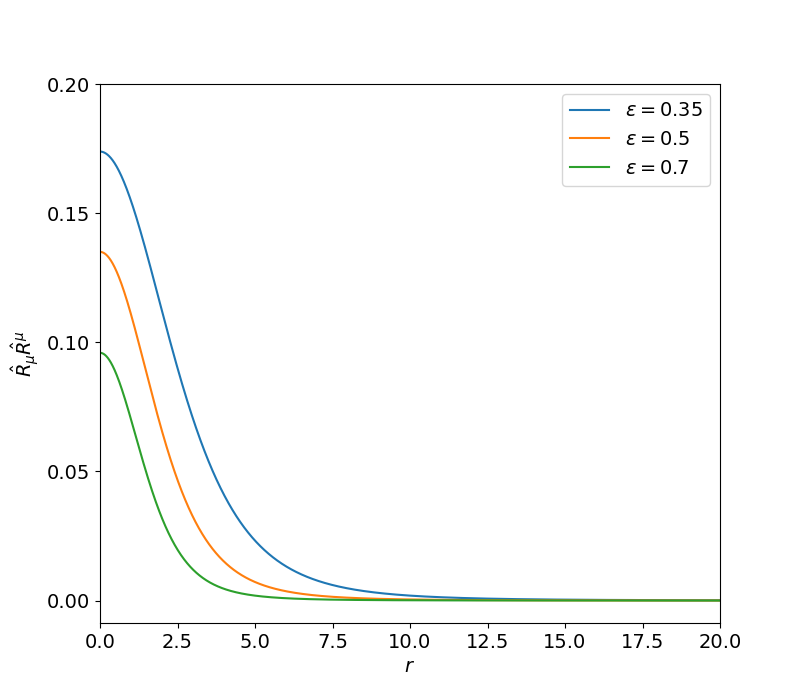}
    }
    \subfloat{
        \centering
        \includegraphics[trim={0cm 0.5cm 1.5cm 1.8cm},clip,width=0.48\linewidth]{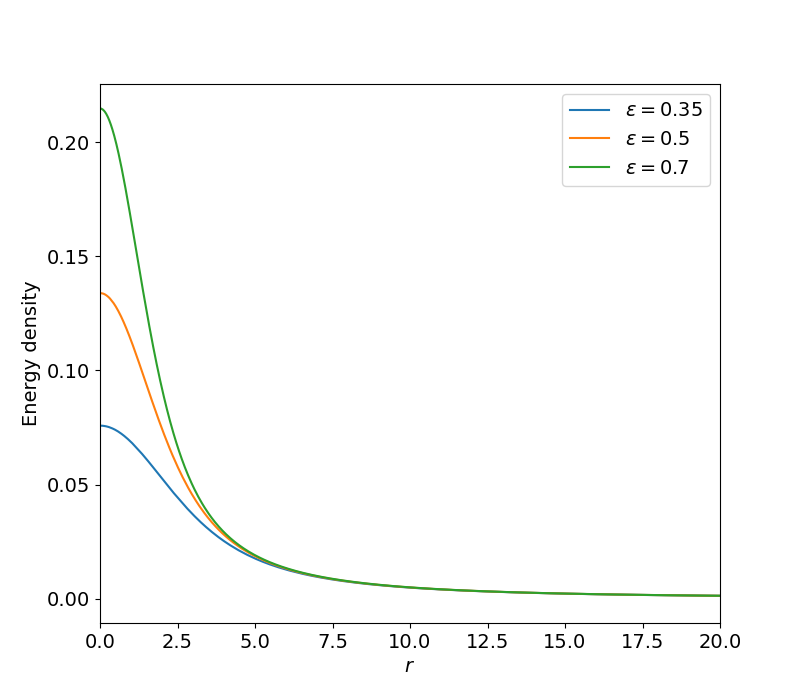}
    }
    \caption{We present the numerical solutions of charge-breaking monopoles, for a range of $\epsilon$, with $f(r)$ displayed in the top left plot, $f_+(r)$ in the top right plot, $\hat{R}_\mu \hat{R}^\mu$ on the bottom left and the energy density on the bottom right. Note that $\hat{R}_\mu = \lambda_1R_\mu/\mu_1^2$ and that the y-axis for the $f(r)$ plot does not start at zero ($f(0)=f_+(0)$).}
    \label{fig: charge-breaking profiles}
\end{figure}

In Figure \ref{fig: charge-breaking profiles} we present the solutions to the above equations (found numerically using a grid of $10000$ points and $\Delta x = 0.01$) as well as $\hat{R}_\mu \hat{R}^\mu$ - where $R_\mu = v_{\rm SM}^2\hat{R}_\mu$ - and the energy density. It is clear that these solutions have $R_\mu R^\mu \neq 0$ close to the monopole core and that its value decreases with increasing $\epsilon$ - a feature that is also present in the 3D simulations. We can also see that the length scale associated with $\epsilon$ corresponds to the length scale of $f_+$, while $f$ changes only marginally. Finally, although the energy of these solutions is infinite because they are global monopoles, the energy density decreases with increasing $\epsilon$, which likely means that the total energy of these monopoles in a more realistic gauged theory will increase with $\epsilon$.

\section{Discussion}
\label{sec:discussion}
    
The 2HDM can predict a variety of topological defects when accidental symmetries are broken. Specifically, for $\text{SU}(2)_L \times \text{U}(1)_Y$ preserving symmetries, there are {\em three} domain wall solutions, {\em two} vortex solutions and {\em one} global monopole solution~\cite{Brawn2011}. Here, we have focused exclusively on the 2HDM with an $\text{SO}(3)_\text{HF}$ symmetry that produces global monopoles.

Our simulations clearly show the formation of monopoles and that they evolve as one might expect with $N_{\rm mono}\propto t^{-3}$ for the values of $\epsilon$ that are easily investigated with our simulations and within the dynamic range. However, the structure of the monopoles formed is very different to the solution expected from ref.~\cite{Brawn2011}. There appears to be a violation of the neutral vacuum in the centre of the monopole with $R^{\mu}R_{\mu}$ and $v_{+}$, and the field configuration has non trivial structures in the vector field $n^{a}$ which we have shown encode the standard model degrees of freedom. A similar phenomenon was found in the case of domain walls~\cite{Viatic2020,Law:2021ing}, albeit in a lower dimensional situation.

We have presented a new monopole ansatz for the 2HDM that is closely related to the Nambu monopole, but it is stable and has no requirement for there to be a string emerging from one of the poles that connects it to an anti-monopole. It is spherically symmetric and has the property that $R^a = n^a \propto \hat{r}^a$ and we have shown that this is also the case for the monopoles formed in our simulations, after a constant rotation is performed. The energy minimising solution under this ansatz always has non-zero $R_\mu R^\mu$ (as $f_+=0$ is not a solution) in the core of the monopole, but it decreases as $\epsilon$ is increased, presumably going to zero in the limit where $\epsilon \to \infty$.

We note as a final point, that topological defects which violate the neutral vacuum conditions in the core and hence become superconducting can have some novel interaction properties with photons and other relativistic particles~\cite{Battye:2021dyq}. Here, we have confirmed that this indeed happens in the 2HDM model with $\text{SO(3)}_{\rm HF}$ symmetry and it is clearly worth exploring the cosmological consequences of this. There are a number of other possibilities, for example, mixing of photons with Z and W bosons, generation of primordial magnetic fields and possibly even the violation of SM constraints on the interaction of leptons and quarks, although this will depend on the structure of the Yukawa sector of the theory which we have not discussed here. Moreover, Nambu monopoles in the SM have been discussed in the context of primordial magnetic fields~\cite{Vachaspati:2020blt,Patel2022}. We have shown that the production of such monopoles maybe much more natural in the context of the 2HDM and this deserves further investigation.

\section*{Acknowledgements}
RB and DV would like to thank Apostolos Pilaftsis for his collaboration on related work on domain walls.

\bibliography{bibliography}
\bibliographystyle{unsrt}

\end{document}